\documentclass{article}
\usepackage{amsmath,graphicx}
\usepackage{xcolor,booktabs,array}
\usepackage{url}
\usepackage{algorithm,algpseudocode}
\usepackage{amsmath}
\usepackage{amsfonts}
\usepackage{amssymb}
\usepackage{enumitem}
\usepackage{INTERSPEECH2020}

\title{Speech Denoising by Accumulating Per-Frequency Modeling Fluctuations}
\name{Michael Michelashvili, Lior Wolf}
\address{
The School of Computer Science\\
  Tel Aviv University}
\email{mosheman5@gmail.com, wolf@cs.tau.ac.il}

\begin{document}

\maketitle
\begin{abstract}
We present a method for audio denoising that combines processing done in both the time domain and the time-frequency domain. Given a noisy audio clip, the method trains a deep neural network to fit this signal. Since the fitting is only partly successful and is able to better capture the underlying clean signal than the noise, the output of the network helps to disentangle the clean audio from the rest of the signal. This is done by accumulating a fitting score per time-frequency bin and applying the time-frequency domain filtering based on the obtained scores. The method is completely unsupervised and only trains on the specific audio clip that is being denoised. Our experiments demonstrate favorable performance in comparison to the literature methods. Our code and samples are available at~\url{github.com/mosheman5/DNP} and as supplementary.
\end{abstract}

\section{Introduction}

Many unsupervised signal denoising methods work in a similar way. First, a spectral mask is estimated, which predicts for every frequency, whether it is relevant to the clean signal or mostly influenced by the noise. Then, one of a few classical methods, such as the Wiener filter~\cite{lim1979wiener} or \textcolor{black}{LSA estimator}~\cite{ephraim1985lsa} are used to clean the audio.



In this work, we train a network to fit the input signal, and observe the part of the signal that has the largest amount of uncertainty, i.e., which was modeled most poorly. Assuming that the clean signal, in the time domain, is well-captured by a deep convolutional neural network, this part is masked out by one of the classical speech-enhancement methods.

Our method achieves results that are comparable to the state-of-the-art unsupervised literature methods and approach those of the supervised methods. Moreover, our method shows robustness and is applicable in scenarios in which the SOTA method fails. Note that similar to most unsupervised methods in the literature, the method observes only the input signal and does not observe other signals in the dataset.

\subsection{Related work}

\noindent{\bf Unsupervised Noise estimation algorithms\quad}
Traditional noise estimation algorithms typically assume that the speech signal contains pauses and low-energy segments where statistics of the noise can be measured, and that the noise is more stationary than the speech signal. The algorithms can be divided into three main categories: minimal-tracking algorithms~\cite{doblinger95, martin2001}, which find the minimum for each frequency bin using a short time window; time-recursive averaging algorithms~\cite{ephraim1985lsa, cohen2002mcra, cohen2003imcra,sørensen2005, rangachari2006}, which average over time in order to provide the noise estimation; and histogram-based algorithms~\cite{hirsch1995}, in which the noise power is assumed to be the most occurring value in the amplitude histogram.

The estimated a-priori SNR of the noise signal is then used as an input for one of a few classical speech enhancement algorithms. These algorithms multiply the original signal with a gain calculated from the a-priori SNR, either directly, as in the Wiener filter, or using a regularization over the time-freqeuncy domain as described in ~\cite{ephraim1985lsa}.

\noindent{\bf Supervised Noise estimation algorithms\quad}
Supervised speech denoising algorithms observe, during training, both the noisy sample and the underlying clean samples and learn to map from noisy samples to clean samples. The SEGAN method~\cite{pascual2017segan} employs an encoder-decoder architecture, which is trained with an additional GAN loss~\cite{gan}. The Wavenet method ~\cite{rethage2018} employs non-causal Wavenet~\cite{wavenet2016} architecture with regression loss rather than discrete softmax output distribution. 
Deep Feature Loss~\cite{geramin2018dfl} method uses a context aggregation network, and instead of using the MSE loss between the output and the target, employs a perceptual loss function. The perceptual loss is derived from the deep layer activations of a network  that is pre-trained for audio classification tasks.
The current state-of-the-art method ~\cite{Koizumi2020} learns auxiliary speaker-aware features during train to improve accuracy. Further more, they use multi head self attention ~\cite{transformer} for better capture of the noise statistics


\noindent{\bf Deep Image Priors\quad} In the DIP method~\cite{dip} a CNN of a given architecture is trained to produce the input image as its output, given a random tensor as the network's input. Based on the assumption that the CNN fits a clean image much faster than it fits a noisy signal, the algorithm is stopped after it starts to fit the given image, but before it fits all of its details. While DIP trains a network to fit the data similarly to our method, we note that the cleaning of audio signals is much more involved. In computer vision, the clean image emerges from the learned network simply as its output. In contrast, applying this idea to audio, the network produces an output that is unacceptable in quality, which means that the DIP method cannot be successfully applied to audio. We also note that while the DIP networks train much faster on a clean image than on a mixed signal that contains both image and noise, in audio, moderate amounts of noise do not significantly change the convergence speed. This means that a DIP prior cannot even readily distinguish if the input singal in audio is noisy or not.

DAP is an audio adapted version of DIP uses harmonic convolutions which take into account the harmonic structure of audio signals~\cite{DAP-2020}. Convincing results are shown on added Gaussian noise but the method is not being applied to non-stationary noise signals, which are the common scenario in speech signal corruption and are the type of signals we explore in this paper.



\section{Method}

We employ the CNN architecture known as the WaveUnet~\cite{stoller2018waveunet}, which consists of an encoder-decoder architecture with skip-connections between the two subnetworks. Assuming an additive noise model, where the clean signal $x$ and the noise $n$ are unknown, we create a random input signal $z$ of the same dimension as the noisy signal $y=x+n$,  and train the network $f=f_\theta$ to fit the noisy signal, i.e., we solve the minimization problem
    $\min_\theta \|f_\theta(z)-y\|$, 
where $\theta$ is the parameter vector (weights and biases) of network $f$.

\begin{figure*}[t]
  \centering
\noindent\begin{tabular}{ccc}
\includegraphics[height=.229303\linewidth,width=.3126302395\linewidth,trim={0 0 0 0},clip]{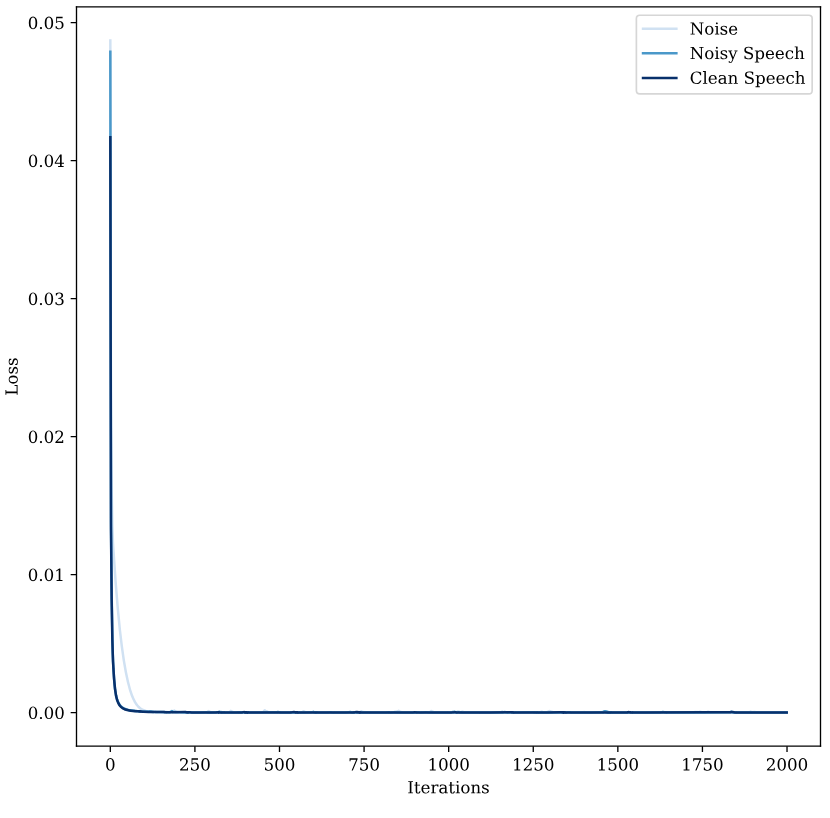} &
\includegraphics[height=.229303\linewidth,width=.3126302395\linewidth,trim={0 0 0 0},clip]{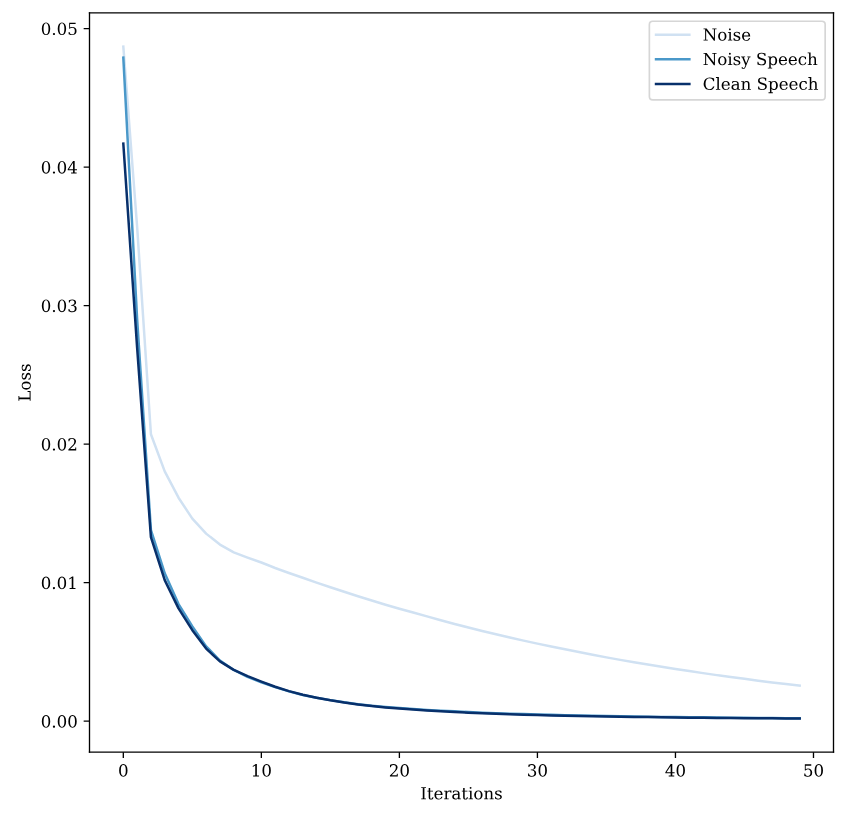} &
\includegraphics[height=.229303\linewidth,width=.3126302395\linewidth,trim={0 0 0 0},clip]{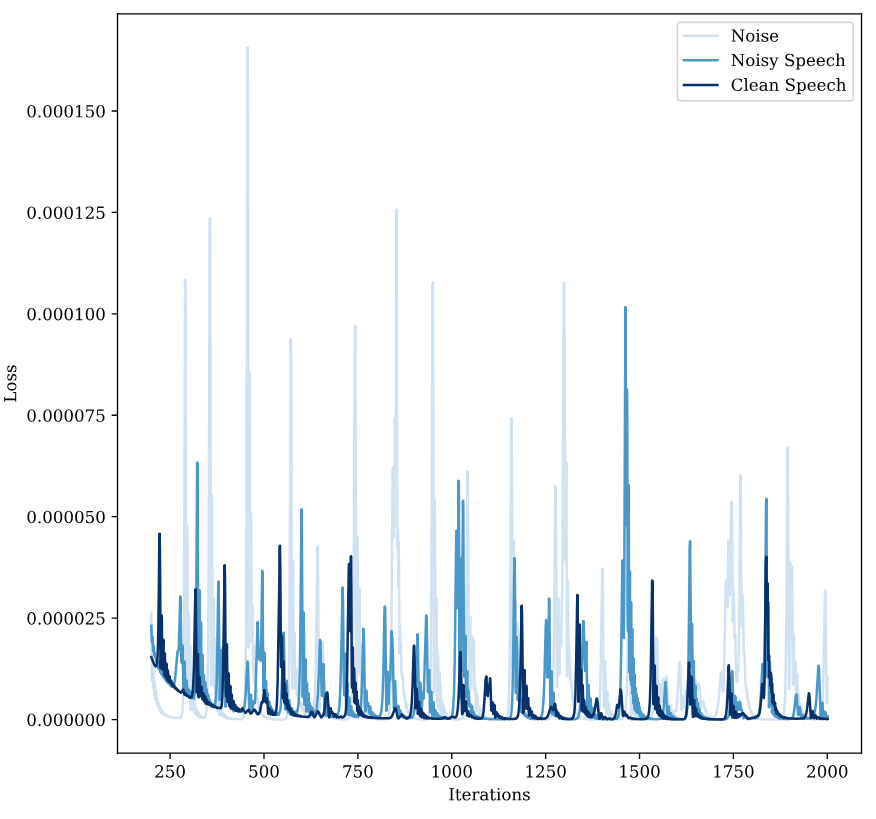} \\
(a) & (b) & (c)\\
 \end{tabular}
 \vspace{-8pt}
  \caption{Typical loss profiles obtained during training for a signal that is clean, noisy, or entirely noise. (a) first 2000 iterations. (b) zoom-in to the first 50 iterations. (c) zoom in to iteration 250 onward.}\label{fig:noiseimp}
\end{figure*}

As can be seen in the example given in Fig.~\ref{fig:noiseimp}, the network fits clean speech or music signals ($x$) much faster than it fits noise signals ($n$). However, unlike the situation in computer vision, there is little difference in the coarse behavior between the clean signal ($x$) and the signal with the added noise ($y$).

\begin{figure*}[t]
  \centering
\newlength\q
\setlength\q{\dimexpr .25\textwidth -2\tabcolsep}
\noindent\begin{tabular}{p{\q}p{\q}p{\q}p{\q}}
 \multicolumn{4}{c}{\includegraphics[width=.995\linewidth,trim={0 350 0 0},clip]{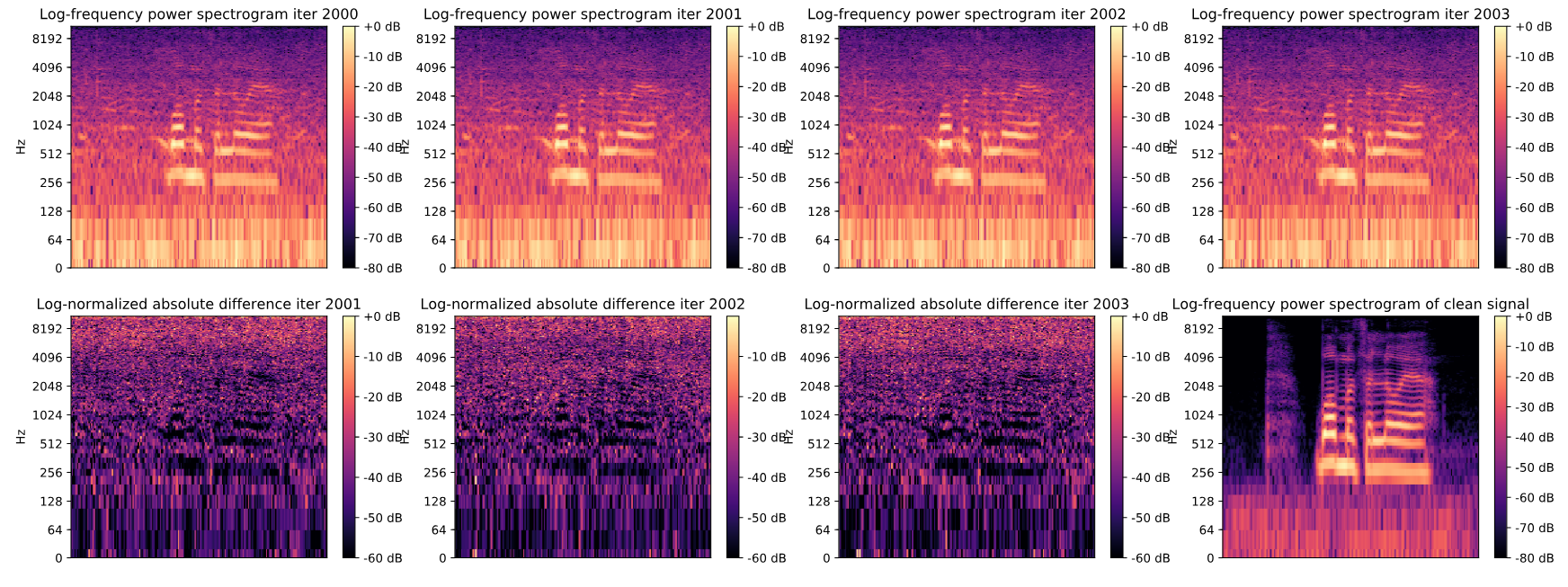}} \\
 \hfil~~~~~~~~(a)& \hfil~~~(b) &\hfil(c)~~~~ &\hfil(d)~~~~~~~~~~ \\
 ~\\
 \multicolumn{4}{c}{\includegraphics[width=.995\linewidth,trim={0 20 0 330},clip]{image.png}} \\
 \hfil~~~~~~~~(e)& \hfil~~~(f) &\hfil(g)~~~~ &\hfil(h)~~~~~~~~~~ \\
 \end{tabular}
 \vspace{-8pt}
  \caption{Instability during training. (a-d) the network output $f_i(z)$ for four consecutive iterations. (e-g) the difference between pairs of consecutive iterations. (h) the spectrogram of the clean signal }\label{fig:duringtraining}
\end{figure*}

While in images, the signal recovered by the network $f_\theta(z)$ during training becomes very similar to $x$, before it starts to resemble $y$, in audio the situation is different. As can be seen in Fig.~\ref{fig:duringtraining}, at every iteration $i$ of training, the current network, which we denote as $f_i$, produces a signal $f_i(z)$ that is only partly denoised. In addition, while in images the method converges to a stable solution, i.e., $f_i(z)\sim f_{i+1}(z)$ after a few training iterations~\cite{dip}, this is not the case in audio. In the case of audio, the network output rapidly changes between iterations. 

Moreover, the noise-free signal $x$ was never reached in our experiments, even after extremely long training sessions. This can also be seen in the baseline experiment we perform (Sec.~\ref{sec:exp}), in which we report the minimal error obtained {when} training the network ($\min_i \|f_i(z)-x\|$). This hindsight experiment, which would have produced a good result in computer vision, produces poor outputs in audio.

\begin{figure}[t]
  \centering
\noindent\begin{tabular}{@{}c@{~}c@{~}c@{~}c@{}}
\includegraphics[width=.48824349232395\linewidth,trim={0 0 0 0},clip]{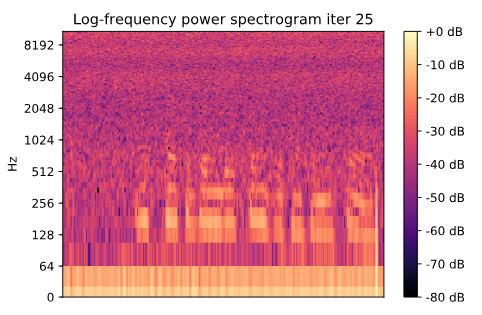} &
\includegraphics[width=.48824349232395\linewidth,trim={0 0 0 0},clip]{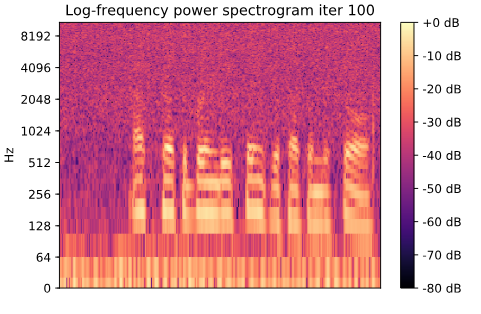} \\
(a) & (b)\\
\includegraphics[width=.48824349232395\linewidth,trim={0 0 0 0},clip]{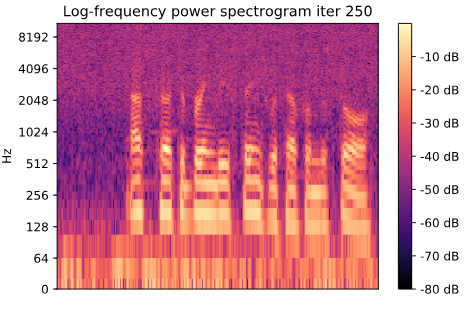} &
\includegraphics[width=.48824349232395\linewidth,trim={0 0 0 0},clip]{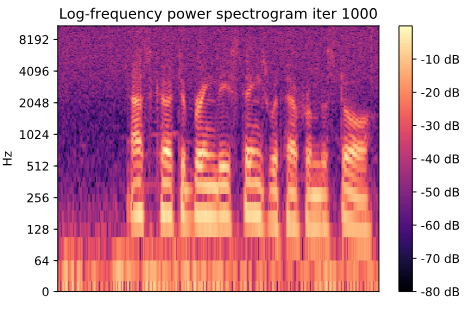}\\
(c) & (d)\\
 \end{tabular}
  \caption{Progress during training. (a) iteration 25. (b) iteration 100. (c) iteration 250. (d) iteration 1000. The low frequencies are learned first, and the higher frequencies follow.}\label{fig:iterations}
\end{figure}

The discussion above does not mean that $f$ does not evolve during training. As can be seen in Fig.~\ref{fig:iterations}, as the iterations progress, the output of $f$ becomes more expressive, and the network models additional frequencies in the signal. 

\begin{figure}[t]
  \centering
\begin{tabular}{cc}
\includegraphics[width=.481682022395\linewidth,trim={0 0 0 0},clip]{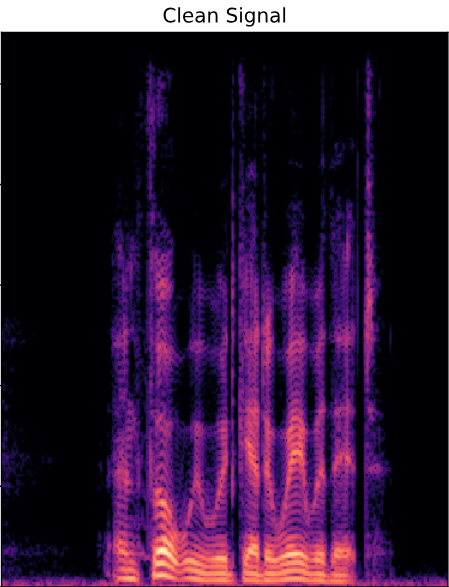} &
\includegraphics[width=.481682022395\linewidth,trim={0 5 0 0},clip]{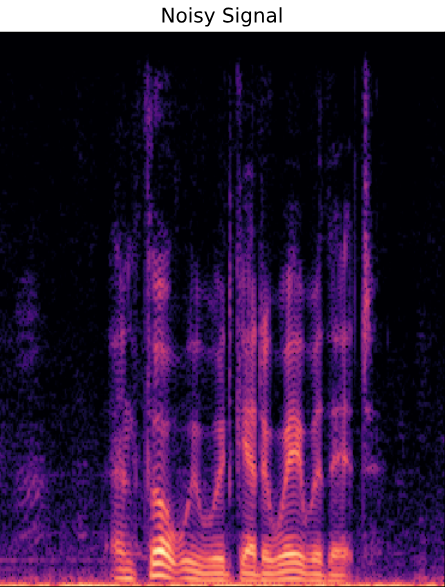} \\
(a) & (b)\\
~\\
\includegraphics[width=.481682022395\linewidth,trim={0 0 0 0},clip]{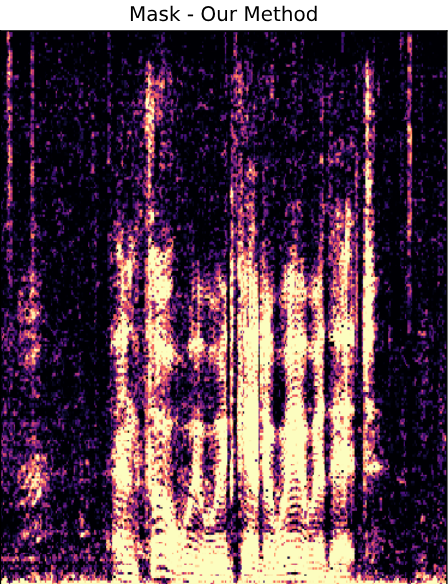} &
\includegraphics[width=.481682022395\linewidth,trim={0 0 0 0},clip]{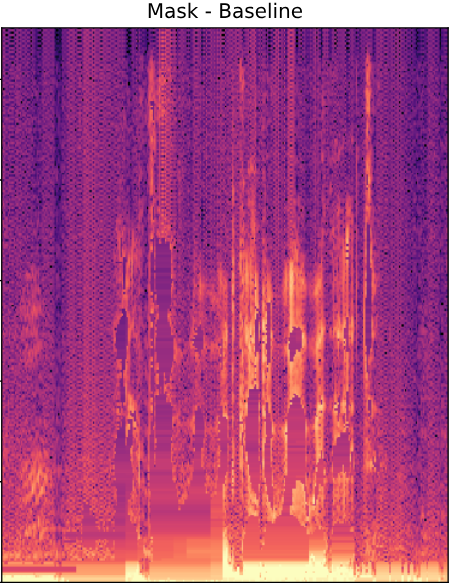} \\
(c) &(d)\\
 \end{tabular}
  \caption{Sample results. (a) The spectrogram of the clean signal $x$. (b) The spectrogram of the noisy signal $y$. (c) The a-priori mask of the signal $M$ returned by our method. (d) The mask obtained by the Connected Frequencies~\cite{sørensen2005} method.}\label{fig:sample}
\end{figure}

\setlength{\textfloatsep}{14pt}

\begin{algorithm}[t]
\caption{The denoising with network priors method}\label{alg:method}
\begin{algorithmic}[1]
\Require{$y$: noisy input,\quad $t$: number of iterations}
\State $\theta_0 \gets \texttt{XavierInit}()$ \Comment{Initialize the weights of $f_0$}
\State $z\sim N(0,1)$ \Comment{Initialize the random vector $z$}
\State ${Y}_0 \gets STFT(f_0(z))$
\State $C=0$
\For{$i\gets1:t$}
\State $\theta_{i} \gets \texttt{semi}\arg\min_\theta \|f_\theta(z)-y\|$ \Comment{One training iteration on $f_{i-1}$, starting with $\theta=\theta_{i-1}$ obtaining $f_{i}$}
\State ${Y}_i \gets \texttt{STFT}(f_i(z))$
\State $H_i \gets (|~|{Y}_i|-|{Y}_{i-1}|~|)/|{Y}_i|$ \Comment{Absolute differences}
\State $P_1 \gets \texttt{percentile}(H_i, 10)$
\State $P_2 \gets \texttt{percentile}(H_i, 90)$
\State $H_i \gets \max(\min(H_i,P_2), P_1)$ \Comment{Clip values}
\State $C \gets C + H_i$ \Comment{Accumulate the differences}
\EndFor
\State $M = (\max(C)-C)/(\max(C)-\min(C))$ \Comment{Normalize}
\State \textbf{return} $M$\Comment{Estimated a-priori SNR of the signal}
\end{algorithmic}
\end{algorithm}

Based on these observations, we propose the method depicted in Alg.~\ref{alg:method} for estimating the a-priori SNR of the clean signal. The input to the method is the signal $y$. Its output is a mask of the dimensions of the STFT, with values in [0,1]. 

After computing a random $z$ vector in line 2, the method undergoes an iterative process for $t$ iterations. Unlike the situation in computer vision, in speech, and other audio signals that we tried, the network $f$ cannot easily fit $y$. Early stopping is, therefore, not a major concern, and we can choose any number of iterations $t$ that is large enough. 

Each iteration consists of the following steps, where $i$ is the iteration index. First, in line 6 of the algorithm, the network $f_{i-1}$ is trained for one iteration, obtaining $f_i$. Then, in line 7, one computes $f_i(z)$ and its STFT $Y_i$. We next compute $H_i$, which is the absolute difference between $|Y_{i-1}|$ and $|Y_{i}|$ normalized by the latter. 

In order to avoid extreme values, every value in $H_i$ that is above the 90th percentile or below the 10th percentile is clipped. An accumulator $C$ sums the resulting matrices (line 12). The accumulator would have high values in the coordinates of the time-frequency domain, in which there is the least stability in the reconstruction of $y$ by the network $f$.

Once the $t$ iterations are over, $C$ is normalized to be in the range of $[0,1]$ (line 14). High accumulated variability implies noise and we, therefore, flip the values, before returning the mask $M$.  With this estimation of the a-pirori SNR, a classical denoising method, such as LSA~\cite{ephraim1985lsa} or the Weiner filter can be used to perform denoising. We employ the former.

\section{Experiments}
\label{sec:exp}

\begin{table}[t]
   \caption{Quantitative evaluation.  Higher scores are better.}
   \label{tab:results}
  \centering
\begin{tabular}{@{}l@{\hspace{-4mm}}c@{~}c@{~}c@{~}c@{~}c@{~}c@{}}
\toprule
Approach & Supervised & CSIG & CBAK & COVL & PESQ & SSNR \\
\midrule
SEGAN~\cite{pascual2017segan} & yes & 3.48 & 2.94 & 2.80 & 2.16 & 7.73\\
{Wavenet}~\cite{rethage2018} & yes & 3.62 & 3.23 & 2.98 & - & -\\
DFL~\cite{geramin2018dfl} & yes & 3.86 & 3.33 & 3.22 & - & -\\
Speaker Aux ~\cite{Koizumi2020} & yes & 4.15 & 3.42 & 3.57 & 2.99 & -\\
\midrule
MCRA~\cite{cohen2002mcra} & no & 2.23 & 2.36 & 1.91 & 1.80 & 5.17\\
IMCRA~\cite{cohen2003imcra} & no & 2.49 & 2.53 & 2.13 & 1.92 & 5.89\\
MCRA2~\cite{rangachari2006} & no & 2.39 & 2.50 & 2.08 & 1.97 & 5.92\\
Martin~\cite{martin2001} & no & 2.48 & 2.61 & 2.21 & 2.12 & 6.37\\
Doblinger~\cite{doblinger95} & no & 2.55 & 2.66 & 2.29 & 2.21 & 6.48\\
Hirsch ~\cite{hirsch1995} & no & 2.67 & 2.66 & 2.35 & 2.21 & 6.26\\
Connected Freq~\cite{sørensen2005} & no & 2.73 & 2.66 & 2.38 & 2.21 & 6.18\\
Wiener~\cite{scalart1996} & no & {\bf 3.23} & 2.68 & {\bf 2.67} & 2.22 & 5.07\\
MMSE-LSA~\cite{ephraim1985lsa} & no & 2.90 & {\bf2.89} & 2.55 & 2.38 & {\bf8.22}\\
DAP~\cite{DAP-2020} & no & 2.50 & 2.62 & 2.18 & 1.99 & 6.33 \\
Ours & no & 3.08 & 2.84 & {\bf 2.67} & {\bf 2.39} & 7.27\\
\bottomrule 
\end{tabular}
\end{table}
\begin{table}[t]
  \caption{Evaluation on samples with trimmed silence.}
  \label{tab:results_trim}
  
  \centering
\begin{tabular}{lc@{~~}c@{~~}c@{~~}c@{~~}c}
\toprule
Approach & CSIG & CBAK & COVL & PESQ & SSNR \\
\midrule
MMSE-LSA~\cite{ephraim1985lsa} & 2.87 & 2.55 & 2.29 & 1.81 & 5.59\\
Our~ & 3.75 & 3.52 & 3.14 & 2.56 & 13.71\\
\bottomrule 
\end{tabular}
\end{table}

\begin{table}[t]
   \caption{Quantitative evaluation with DIP-like audio alternatives.  Higher scores are better.}
   \label{tab:resultsdip}
  \centering
\begin{tabular}{@{}l@{\hspace{-4mm}}c@{~}c@{~}c@{~}c@{~}c@{~}c@{}}
\toprule
Approach & Supervised & CSIG & CBAK & COVL & PESQ & SSNR \\
\midrule
Ours & no & 3.08 & 2.84 & 2.67 & 2.39 & 7.27\\
\midrule
Best $f_i(z)$ & hindsight & 3.04 & 2.36 & 2.39 & 1.81 & 1.59\\
Averaged $f_i(z)$ & no & 3.13 & 2.39 & 2.47 & 1.87 & 1.66\\
\midrule
Noisy Samples & no & 3.35 & 2.44 & 2.63 & 1.97 & 1.68\\
\bottomrule 
\end{tabular}
\end{table}
When applying our method, we employ a WaveUnet with six layers and 60 filters per layer. Each mask filter was produced after $t=5000$ iterations using the Adam optimizer with a learning rate of 0.0005. The method seems insensitive to either of these parameters. We used the  VoiceBank-DEMAND dataset proposed by the authors of~\cite{valentini2017}. 

For the purpose of our work, only the test set has been used. This test set is composed by mixing multiple speakers with 5 different noise types and 4 different SNR setting (2.5, 7.5, 12.5 and 17.5 dB). The original 48 kHz files were downsampled to 16 kHz, the same as other baseline methods~\cite{pascual2017segan, rethage2018, geramin2018dfl, Koizumi2020}. The spectrograms are obtained by using 32 ms Hann window and 8 ms hop length.

For the purpose of computing the results of the baseline unsupervised denoising algorithms, except DAP, the open source metrics evaluation and noise estimation  toolbox~\cite{speechbook}  was used. 

The DAP baseline is our implementation of the method described in~\cite{DAP-2020}. We used a dilations only over the frequency dimension as approximation to DAP algorithm with anchoring of 1.
The scores of the supervised baselines were taken from the respective papers~\cite{pascual2017segan, rethage2018, geramin2018dfl, Koizumi2020}. 

Multiple quality scores are used in order to measure the success of the methods, including (1) CSIG: Mean opinion score (MOS) predictor of signal distortion~\cite{metricsyi2007}, (2) CBAK: MOS predictor of background-noise intrusiveness~\cite{metricsyi2007}, (3) COVL: MOS predictor of overall signal quality~\cite{metricsyi2007}, (4) PESQ: Perceptual evaluation of speech quality ~\cite{pesq2007}, and (5) SSNR: Segmental SNR~\cite{ssnr1988}. All unsupervised methods were post-processed by a highpass filter with a cutoff frequency of 60 Hz, to remove noise below the human speech base frequency. The various measures were computed by using the open source toolbox mentioned above.
A sample result is given in Fig.~\ref{fig:sample}, and the scores are reported in Tab.~\ref{tab:results}. As can be seen from the table, our method outperforms most of the unsupervised literature methods 
except MMSE-LSA~\cite{ephraim1985lsa} in all metrics, with the exception of the CSIG metric, in which it is the second highest method.
MMSE-LSA~\cite{ephraim1985lsa} presents the best SSNR and CBAK results among the unsupervsied methods. The method relies on the unvoiced parts in the beginning of a noisy signal to evaluate the statistics of the noise signal. In Tab.~\ref{tab:results_trim} we present the results of performing denoising on noisy speech clips after trimming the silence at both edges of the clip. As can be seen in the table, in such scenario our method outperforms MMSE-LSA~\cite{ephraim1985lsa}. This experiment shows the robustness of our method. As described in Sec.~2, the noise estimation phase in unsupervised speech denoising algorithms relies on assumptions such as noise statistics or unvoiced segments. When the assumptions are not fulfilled the performance of those algorithms deteriorate. Our model, which does not rely on such heuristics is therefore resilient in various setups.
\begin{figure}[t]
  \centering
\begin{tabular}{@{}c@{~~}c@{}}
\includegraphics[width=.4981682022395\linewidth,trim={0 0 0 0},clip]{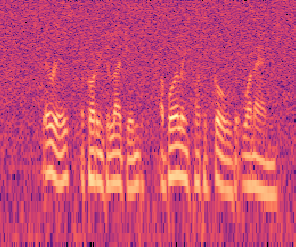} &
\includegraphics[width=.4981682022395\linewidth,trim={0 5 0 0},clip]{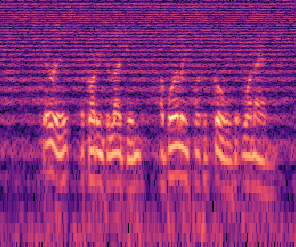} \\
(a) & (b)\\
~\\
\includegraphics[width=.4981682022395\linewidth,trim={0 0 0 0},clip]{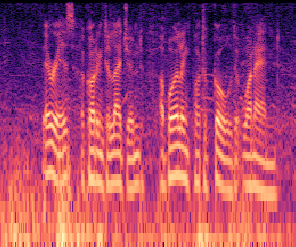} &
\includegraphics[width=.4981682022395\linewidth,trim={0 0 0 0},clip]{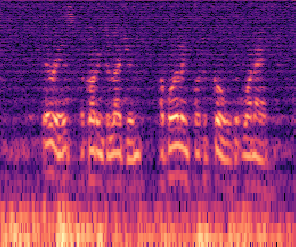} \\
(c) &(d)\\
 \end{tabular}
  \caption{DAP case study. (a) Spectrogram of speech signal contaminated with gaussian noise, SSNR=-5.18 (b) Spectrogram of DAP output after 150 iterations on gaussian contamination, SSNR=-4.31 (c) Spectrogram of speech signal contaminated with non-stationary noise, SSNR=-4.93 (d) Spectrogram of DAP output after 150 iterations on non-stationary contamination, SSNR=-5.02.}\label{fig:dap}
\end{figure}
Our method is also largely comparable to the {\em supervised} SEGAN method~\cite{pascual2017segan}, despite not being trained on any sample outside the single sample $y$, while SEGAN is fully supervised. Our unsupervised method is, however, outperformed by the supervised Deep Feature Loss method~\cite{geramin2018dfl}, which enjoys both a large fully supervised training set and a strong pretrained perceptual loss. The supervised method ~\cite{Koizumi2020} using speaker auxiliary network and a transformer surpassed all methods by large margin. Using self-attention results in efficient noise and speech modeling over time and the speaker auxiliary network applies a strong regularization over the output audio quality.
As shown in Tab.~\ref{tab:results}, our method outperforms DAP~\cite{DAP-2020} by a large margin. Closer inspection raised the fact that using harmonic based convolutions removed additive noise from low SNR signals, but introduced a Gaussian-like noise when trying to model a signal with medium to high SNR, as can be seen in Fig.~\ref{fig:dap}, hence deteriorating many speech signals in the testset.  

To directly compare with DIP~\cite{dip}, we consider the best network output obtained during training $f_i(z)$, $i=1..t$. This happens in hindsight, by comparing it to the clean signal $x$, i.e., it serves as an upper bound to the DIP performance. As can be seen in Tab.~\ref{tab:resultsdip} the statistics for this are very similar to those of the original noisy signal. Averaging multiple network reconstructions $f_i(z)$, $i=250,500,750,\dots,5000$, does not lead to an acceptable result. This is in contrast to the application of DIP in computer vision~\cite{dip}, where the network produces clean outputs during its training and averaging improves results further.

\section{Conclusions}


The advent of deep learning has led to effective supervised denoising algorithms. However, as far as we can ascertain, little progress has been made with deep methods in the unsupervised domain. We explore the usage of model fluctuations based on the unstable behavior of neural audio model while fitting a noisy sample. The method we develop based on this observation is shown to approach the quality of some of the supervised  deep methods. Interestingly, an unsupervised method published over 30 years ago outperforms the other unsupervised methods in the benchmarks. Unlike our method, it assumes the existence of unvoiced samples at the clip's beginning. 
As future work, we would like to explicitly model unvoiced locations,  
but without making this restrictive assumption on the structure of the  signal.

\clearpage
\bibliographystyle{IEEEtran}
\bibliography{audio}

\end{document}